\documentclass[prl,twocolumn,showpacs,preprintnumbers,amsmath,amssymb,superscriptaddress]{revtex4}
\usepackage{epsf}
\usepackage{graphicx}
\usepackage{bm}
\usepackage{amsmath}
\usepackage{color}
\DeclareMathOperator{\arccot}{arccot}
\begin{document}

\title{Resonant exchange interaction in semiconductors}

\author{I.~V.~Rozhansky}
\email{rozhansky@gmail.com} \affiliation{A.F.~Ioffe Physical
Technical Institute, Russian Academy of Sciences, 194021
St.Petersburg, Russia} \affiliation{Lappeenranta University of
Technology, P.O. Box 20, FI-53851, Lappeenranta, Finland}
\author{I.~V.~Kraynov}
\affiliation{A.F.~Ioffe Physical Technical Institute, Russian
Academy of Sciences, 194021 St.Petersburg, Russia}
\author{N.~S.~Averkiev}
\affiliation{A.F.~Ioffe Physical Technical Institute, Russian
Academy of Sciences, 194021 St.Petersburg, Russia}
\author{E.~L\"a hderanta}
\affiliation{Lappeenranta University of Technology, P.O. Box 20,
FI-53851, Lappeenranta, Finland}


\date{\today}

\begin{abstract}
We present a non-perturbative calculation of indirect exchange
interaction between two paramagnetic impurities via 2D free carriers
gas separated by a tunnel barrier. The new method accounts for the
impurity attractive potential producing a bound state. The
calculations show that for if the bound impurity state energy lies
within the energy range occupied by the free 2D carriers the
indirect exchange interaction is strongly enhanced due to resonant
tunneling and exceeds by a few orders of magnitude what one would
expect from the conventional RKKY approach.
\end{abstract}

\maketitle
Semiconductor heterostructures with paramagnetic impurities
spatially separated from the free charge carriers are coming into
focus of the semiconductor-based spintronics. A number of recent
experiments show that paramagnetic ions located at a tunnel distance
from the quantum well (QW) induce substantial spin polarization of
the 2D carriers in the QW while preserving their high
mobility\cite{KorenevSapega,Zaitsev}. Charge carriers tunneling
between the bound impurity states and the continuum of delocalized
states might also play a most important role in the interaction
between the paramagnetic ions themselves. The indirect exchange
interaction between Mn ions mediated by the holes is believed to be
the key mechanism underlying the ferromagnetic ordering in the
InGaAs-based semiconductors doped with Mn\cite{Jungwirth,Aronzon}.
The indirect exchange interaction is usually described on the basis
of RKKY theory which utilizes the second-order perturbation
calculation with account for the Pauli exclusion
principle\cite{Kittel}. The RKKY theory while perfectly applicable
in many cases ignores the fact that attracting potential of the ion
may have a bound state so that the scattering of the free carriers
can be of a resonant character, at that the perturbation theory
fails. In this Letter we report on a new approach to the indirect
exchange pair interaction which takes into account the resonance
case in a non-perturbative way. The exactly solvable Fano-Anderson
model is exploited to describe the tunnel coupling of the bound
state with the continuum\cite{Fano,OURPRB} with the spin
configuration of the impurities being a parameter.

In order to rely on a certain model we consider a heterostructure
containing a QW and $\delta$ layer of paramagnetic ions separated
from the QW by a tunnel barrier. A paramagnetic ion is assumed to
have a bound state characterized by its energy $\varepsilon_0$ while
the QW has a continuum of 2D states starting from the single size
quantization level and filled up to the Fermi level $E_F$. The
resonant condition implies that $\varepsilon_0$ lies within the
energy range of the continuum. Let us consider two parmagnetic ions
located far enough from each other so that the bound state
wavefunctions do not overlap. Both ions are located close to the QW
so that the weak tunneling is allowed. The exchange interaction is
described by:
\begin{equation}
\label{eqHamiltonianExchange}
 H_{J} = {J\widehat{\bf{S}} \left[ {\delta
\left( {{\bf{r}} - {\bf{R}}_1 } \right)\widehat{\bf{I}}_1  + \delta
\left( {{\bf{r}} - {\bf{R}}_2 } \right)\widehat{\bf{I}}_2 }
\right]}.
\end{equation}
where $\bf{R}_{1,2}$ -- the ions positions, $\widehat{\bf{I}}_{1,2},
\widehat{\bf{S}}$ -- the spin operators for the ion and the free
carrier respectively, $J$ -- the exchange constant.
The RKKY theory\cite{Kittel} gives the interaction energy
proportional to $J^2<\widehat{\bf{I}}_1\widehat{\bf{I}}_2>$. Since
our theory also does not produce any terms linear in $J$ we can from
the very beginning replace the ions spin operators by the classical
moments $I_1$,$I_2$ and treat them as parameters. The indirect
exchange energy can be then evaluated as the energy difference
between parallel and antiparallel spin configurations of the two
impurity ions. For the (anti)parallel ion spin configuration $H_J$
(\ref{eqHamiltonianExchange}) does not mix the free carrier spin
projections so we can replace $\widehat{\mathbf{S}}$ with a
parameter $s=\pm|s|$.
The total Hamiltonian
consists of three terms:
\begin{equation}
\label{eqH} H = H_0  + H_T  + H_J, \end{equation} where $H_0$ -- the
Hamiltonian of the system without tunnel coupling and spin-spin
interaction, $H_T$ -- the Bardeen's tunnel term\cite{Bardeen}, $H_J$
-- the exchange interaction term (\ref{eqHamiltonianExchange}). In
the second quantization representation:
\begin{align}
 H_0  &=
\varepsilon _0 a_1 ^ +  a_1  + \varepsilon _0 a_2 ^ +  a_2  + \int
{\varepsilon _\lambda  c_{\lambda}^ +  c_{\lambda } } d\lambda,
  \nonumber\\
 H_T  &=
\int {\left( {t_{1\lambda } c_{\lambda }^ +  a_1  + t_{2\lambda }
c_{\lambda }^ +  a_2  + h.c.} \right)d\lambda },
  \nonumber \\
 H_J  &= JA\left( {I_1 sa_1^ +  a_1  + I_2 sa_2^ +  a_2 } \right) ,
  \end{align}
where $a^+_{1,2},a_{1,2}$ -- the creation and annihilation operators
for the bound states at the impurity ions $1,2$, characterized by
the same energy $\varepsilon_0$ and localized wavefunctions
$\psi_1$, $\psi_2$. $c^+_{\lambda}, c_{\lambda}$ -- the creation and
annihilation operators for a continuum state characterized by the
quantum number(s) $\lambda$, having the energy $\varepsilon
_{\lambda}$ and the wavefunction $\varphi_{\lambda}$, energy is
measured from the QW size quantization level,
\begin{equation}
\label{eqA} A = \left| {\psi _1 \left( {{\bf{R}}_{\bf{1}} } \right)}
\right|^2 = \left| {\psi _2 \left( {{\bf{R}}_2 } \right)} \right|^2.
\end{equation}
The tunnel parameters are given by\cite{Bardeen,OURPRB}:
\begin{equation}
\label{eqT} t_{1,2} \left( \lambda  \right) =  - \frac{{\hbar ^2
}}{{2m_{\perp}}}\int_{\Omega _S } {dS} \left( {\varphi _{_\lambda  }
^* \frac{d}{{dz}}\psi _{1,2}  - \psi _{1,2} \frac{d}{{dz}}\varphi
_\lambda } \right),
\end{equation}
where integration is over the plane $\Omega_S$, parallel to the QW
plane and passing through the ions centers, $m_\perp$ is the
effective mass in the direction perpendicular to the QW plane.
 The hybridized eigenfunctions $\Psi$ of the whole system
 can be expanded over the bound states and the delocalized
states in the form:
\begin{equation} \label{eqExpand}
\Psi  = \nu _1 \psi _1  + \nu _2 \psi _2  + \Phi,\,\,\,\,\;\Phi=\int
{\nu _{_\lambda } } \varphi _\lambda  d\lambda.
\end{equation}
 Plugging
(\ref{eqExpand}) into the stationary Schrodinger equation
$H\Psi=E\Psi$ with Hamiltonian (\ref{eqH}) yields:
 \begin{align}
 \label{eqFanoSystemContinous}
 \nu _1 \left( {\varepsilon _0  - E + JAI_1 s} \right) + \int {t_{1\lambda } \nu _\lambda  d\lambda }  = 0 & \nonumber\\
 \nu _2 \left( {\varepsilon _0  - E + JAI_2 s} \right) + \int {t_{2\lambda } \nu _\lambda  d\lambda }  = 0 & \nonumber\\
 \nu _\lambda  \left( {\varepsilon  - E} \right) + \nu _1 t_{1\lambda } ^*  + \nu _2 t_{2\lambda } ^* = 0 &.
 \end{align}
According to the Fano method\cite{Fano} $\nu_{\lambda}$ is expressed
from the last equation of (\ref{eqFanoSystemContinous}) as follows:
\begin{equation}
\label{eqnualpha} \nu _\lambda   = P\frac{{\nu _1 t_{1\lambda } ^* +
\nu _2 t_{2\lambda } ^* }}{{E - \varepsilon }} + Z\left( {\nu _1
t_{1\lambda } ^*  + \nu _2 t_{2\lambda } ^* } \right)\delta \left(
{E - \varepsilon } \right)
 ,
\end{equation}
where $P$ denotes principal value and $Z(E)$ is to be determined.
Plugging (\ref{eqnualpha}) into (\ref{eqFanoSystemContinous})
yields:
\begin{align}
\label{eqSystemnu}
& \nu _1 \left( { JAI_1 s + F_{11}  + ZT_{11}-E' } \right) + \nu _2 \left( {F_{21}  + ZT_{21} } \right) = 0 \nonumber \\
& \nu _1 \left( {F_{12}  + ZT_{12} } \right) + \nu _2 \left( {
JAI_2s + F_{22}  + ZT_{22}-E' } \right) = 0,
\end{align}
where
\begin{align}
\label{eqFT} &F_{\alpha \beta }  = P\int  {\frac{{t_{\alpha \lambda
} ^* t_{\beta \lambda } }}{{E - \varepsilon }}d\lambda },\;\;\;
T_{\alpha \beta }  = \int {t_{\alpha \lambda } ^* t_{\beta \lambda }
\delta \left( {\varepsilon  - E} \right)d\lambda },\nonumber\\
&E'=E-\varepsilon_0,\;\;\;\alpha,\beta=1,2.
\end{align}
For non-trivial solution of (\ref{eqSystemnu}) one gets a dispersion
equation for $Z$, which determines the energy-dependent phase shift
due to the scattering at the bound state\cite{Fano}. The phase
shifts affect the density of the delocalized states and, in this
way, the whole energy of the system with the fixed number of the
free carriers. Since the phase shifts are different for the parallel
and antiparallel ions spins configurations so is the total energy.
This difference is interpreted as the indirect exchange interaction
energy.

To proceed to the specific case let us consider two ions located at
the same distance $d$ from the QW having the distance $R$ between
them. The z-axis is normal to the QW plane ($z=0$ corresponds to the
QW boundary), x-axis passes through the ions centers with $x=0$ in
the middle of them. Thus, the coordinates of the ions are:
\[
{\bf{R}}_1  = \left( { - R/2,0,d} \right);\,\,\,{\bf{R}}_2  = \left(
{R/2,0,d} \right).
\]
Because it is assumed $R>>d$ and the localized wavefunctions
$\psi_1,\psi_2$ do not overlap, their particular form is not
important. It is convenient to take the localized wavefunctions in
the form:
\begin{equation}
\label{eqLocalized} \psi _{1,2}  = \left( {\frac{2}{{\pi r_0 ^2 }}}
\right)^{3/4} e^{ - \left( {\frac{{x \pm R/2}}{{r_0 }}} \right)^2 }
e^{ - \left( {\frac{y}{{r_0 }}} \right)^2 } e^{ - \left( {\frac{{z -
d}}{{r_0 }}} \right)^2 },
\end{equation}
where $r_0$ is the localization radius. The continuum wavefunctions
are taken as follows:
\begin{equation}
\label{eqWavefunCont} {\varphi _{\bf{k}}} = \eta \left( z
\right){e^{i{\bf{k\rho}}}} \end{equation} Here $\mathbf{k}$ is the
in-plane wavevector, $\mathbf{\rho}$ -- 2D in-plane radius-vector,
$\eta \left( z \right)$ is the envelope function of size
quantization along $z$. Outside of the QW:
\begin{equation}
\eta \left( z \right) = \zeta{{}}{a^{-1/2}e^{ - qz}},
\end{equation}
where $q=\sqrt{{{2m_{\perp}{E_0}}}}/{{{\hbar ^2}}}$, $E_0$ is the
binding energy of the bound state, which at the same time determines
the height of the potential barrier between impurities and the
QW\cite{OURLowTempPhys}, $a$ is the QW width, $\zeta$ is a
dimensionless parameter weakly depending on $q$ and $a$. For a
realistic rectangular QW $\zeta\approx0.5$. The calculation of
(\ref{eqT}) using (\ref{eqLocalized}) (assuming $r_0\ll k^{-1}$) and
(\ref{eqWavefunCont}) yields :
\begin{equation}
\label{eqtunpar} t_{1,2} \left( k \right) = \sqrt{\frac{{\hbar ^2 T
}}{{2\pi m}}}e^{ - i{\bf{kR}}_{{\bf{1,2}}} } ,
\end{equation}
where $T$ -- the energy parameter for the tunneling:
\begin{equation}
T =  {\left( {2\pi } \right)^{3/2} \zeta ^2 \frac{{r_0 m}}{{am_ \bot
}}} E_0 e^{ - 2qd},
\end{equation}
$m$ -- the effective mass along the QW plane. Plugging
(\ref{eqtunpar}) into (\ref{eqFT}) we get:
\begin{equation}
\label{eqFmany}
\begin{array}{l}
 T_{11}  = T_{12}  = T, \;\; T_{12}  = T_{21}  \equiv t = TJ_0 \left( {kR} \right), \\
  F_{11}=F_{22}\equiv F,\;\; F_{12}  = F_{21}  \equiv f = \pi TY_0 \left( {kR} \right),
 \end{array}
\end{equation}
where $J_0, Y_0$ -- Bessel and Neumann functions of zeroth order,
$k=\sqrt{2mE}/\hbar$. The quantity $F$ represents the shift of the
resonance position with respect to $\varepsilon_0$, its explicit
calculation requires more accurate expression than (\ref{eqtunpar})
taking into account $k\sim r_0^{-1}$ to avoid the divergence.
However, it will be not needed since  $F$ is of the order of $T$ and
does not depend on R.
From (\ref{eqSystemnu}) follows the dispersion equation for $Z$:
\begin{align}
\label{eqDisp} &\left({ZT+F - E' + JAI_1 s} \right)\left( {ZT +F- E'
+ JAI_2 s} \right) = \nonumber\\
&=\left( {f + Zt} \right)^2.
\end{align}
 For the parallel spin configuration $I_1=I_2=I$ the two roots
are:
\begin{align}
\label{eqZparallel}
 Z_\pm  = -\frac{{JAI s +F \pm f  - E'}}{{T  \pm t }}.
\end{align}
$Z_\pm$ corresponds to $\nu_1=\pm\nu_2$ so that the hybridized
wavefunction (\ref{eqExpand}) is either symmetric or antisymmetric
with respect to $x\rightarrow-x$. This is due to the symmetry of the
spin-spin interaction which holds only for the parallel spin
configuration. With use of (\ref{eqExpand}) and (\ref{eqnualpha})
the delocalized part of the hybridized wavefunction is given by:
\begin{align}
\label{eqPhiplusminus} \Phi _ \pm   = C_ \pm  \left[
\begin{array}{l}
 J_0 \left( {k\rho _1 } \right)\cos \Delta _ \pm   - H_0 \left( {k\rho _1 } \right)\sin \Delta _ \pm   \\
  \pm J_0 \left( {k\rho _2 } \right)\cos \Delta _ \pm   \mp H_0 \left( {k\rho _2 } \right)\sin \Delta _ \pm   \\
 \end{array} \right],
\end{align}
where $J_0$ and $H_0$ are Bessel and Struve functions of the zeroth
order, $\rho_{1,2}=|\mathbf{\rho}-\bf{R_{1,2}}|$,
\[
\tan \Delta_\pm\left(E\right)  =  - \frac{\pi }{{Z_\pm\left( E
\right)}},\,\,\,\,\,C_ \pm   =  - \frac{{\pi T\nu _1 \eta \left( d
\right)}}{{\sin \Delta _ \pm  }}.
\]
The general solution is an arbitrary linear combination:
\[
\Phi \left( {\bf{\rho }} \right) = A\Phi _ +  \left( {\bf{\rho }}
\right) + B\Phi _ -  \left( {\bf{\rho }} \right).
\]
 Let us put the system in a big cylindrical box of radius
$L$ and apply the boundary conditions $\Phi(\rho=L)=0$ (and
independent on the polar angle). Using the asymptotic forms of
$\Phi_+$,$\Phi_-$ for large $L$ we obtain the two solutions:
\begin{align}
& B=0,\, \Phi _ +  \left( L \right) = 0\, \to \,\cot \left( {k L - \pi /4} \right) = \tan \Delta_+ , \nonumber \\
&A=0,\,\Phi _ -  \left( L \right) = 0\, \to \,\tan \left( {k L - \pi
/4} \right) =  - \tan \Delta_-.
\end{align}
This gives the following quantization condition for k:
\begin{equation}
 k_\pm = {k_L} - \frac{{{{\widetilde{\Delta }_\pm}}}}{L},
 \end{equation}
 where
 \begin{align}
 \widetilde{\Delta} _ +  &=  - \arccot \left[ {\frac{{\pi \left( {T+t } \right)}}{{JAI s + F- E'+f  }}} \right], \nonumber \\
 \widetilde{\Delta} _ -  &= \arctan \left[ {\frac{{\pi \left( {T-t } \right)}}{{JAI s + F- E'-f }}}
 \right],
 \end{align}
$k_L=\pi n/L,\;n=1,2,3,...$ -- the quantized wavenumber in the
abscence of the tunnel coupling with the localzied states. For the
discrete energy levels in a box we have\cite{Muller}:
\begin{equation}
\label{eqDiscreteEnergyDiff} \varepsilon _\pm = {\varepsilon _L} -
\frac{{{\hbar ^2}{k_L}{\widetilde{\Delta} _\pm}\left( {{\varepsilon
_L}} \right)}}{{mL}} + O\left( {\frac{1}{{{L^2}}}}
\right),\end{equation} where ${\varepsilon _L}=\hbar^2k_L^2/2m$.

\begin{figure}
  \leavevmode
 \centering\includegraphics[width=0.48\textwidth]{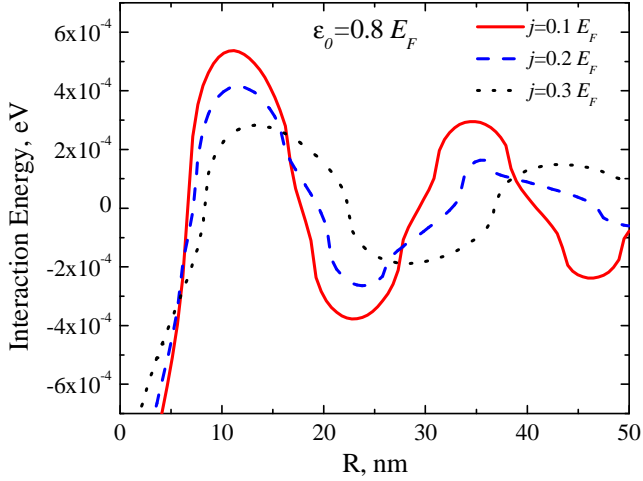}
 \caption{(Color online) Indirect exchange interaction energy vs distance between
 ions in the resonant case}
 \label{figResonance08Ef}
\end{figure}

\begin{figure}
  \leavevmode
 \centering\includegraphics[width=0.48\textwidth]{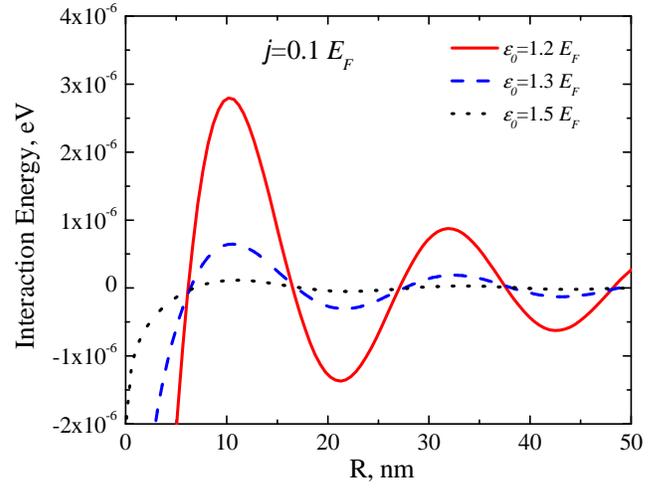}
 \caption{(Color online) Indirect exchange interaction energy vs distance between
 ions
 in the non-resonant case}
 \label{figNonRes}
\end{figure}

Let us now consider the antiparallel configuration of the ions spins
 $I_1=-I_2=I$. The dispersion equation (\ref{eqDisp}) again has two roots $Z_1$,
 $Z_2$,
but unlike the previous case the corresponding wavefunctions
$\Phi_{1,2}$ are neither symmetric nor antisymmetric, they can be
represented as a superposition of symmetric and antisymmetric parts:
\begin{equation}
\Phi _{1,2}  = c_ +  \left( {Z_{1,2} } \right)\Phi _ +  \left(
{Z_{1,2} } \right) + c_ -  \left( {Z_{1,2} } \right)\Phi _ -  \left(
{Z_{1,2} } \right),
\end{equation}
where
\[
\frac{{c_ +  }(Z)}{{c_ -  }(Z)} = \frac{{F- E'- f   + Z\left( {T - t
} \right) + JAIs}}{{F - E'+f   + Z\left( {T + t } \right) + JAIs}},
\]
and $\Phi_+$, $\Phi_-$ are given by (\ref{eqPhiplusminus}). The
general solution is a linear combination of $\Phi_1$ and $\Phi_2$.
The quantization in a finite size box results in:
\begin{align}
 \label{eqAntiDelta}
& k_{1,2} = {k_L} - \frac{{{{\Delta }_{1,2}}}}{L}, \nonumber\\
& {{\Delta }_1} = \arctan \left[ {\frac{{\pi \left( {F- E'+f  } \right)\left( {{T} - {t}} \right)}}{{{{\left( {{F} - E'} \right)}^2} - {f}^2 - {{\left( {JAIs} \right)}^2}}}} \right], \nonumber\\
& {{\Delta }_2} =  - \arccot \left[ {\frac{{\pi \left( {F- E'-f  }
\right)\left( {{T} + {t}} \right)}}{{{{\left( {{F} - E'} \right)}^2}
- {f}^2 - {{\left( {JAIs} \right)}^2}}}} \right].
 \end{align}
Given the discrete energy levels for the parallel and antiparallel
ions spin configurations the indirect exchange energy can be
calculated by summing the energy difference over all free carriers.
Using (\ref{eqDiscreteEnergyDiff}) we have:
\[
E_{exc} =  - \frac{1}{\pi }\sum\limits_s {\int\limits_0^{E_F }
{\left[ {\left( {\widetilde \Delta _ +   + \widetilde \Delta _ -  }
\right) - \left( {\Delta _1  + \Delta _2 } \right)} \right]dE} }.
\]
 The evaluation neglecting terms of the order higher than
$T^2$ yields:
\begin{align}
\label{eqExchangeFinal} E_{exc} = \frac{1}{\pi }\int\limits_0^{E_F }
{\arctan \left[ {\frac{{8\pi ^2 T^2 j^2 J_0 \left( {kR} \right)Y_0
\left( {kR} \right)}}{{\left( {\left( {\varepsilon  - \varepsilon _0
} \right)^2  - j^2 } \right)^2 }}} \right]d\varepsilon }
 ,
\end{align}
where $k=\sqrt{2m\varepsilon}/\hbar$, $j=|JAIs|$. As seen from
(\ref{eqExchangeFinal}) the interaction energy $E_{exc}$ oscillates
with the distance between the impurities $R$. The argument of
arctangent in (\ref{eqExchangeFinal}) has poles at
$\varepsilon=\varepsilon_0\pm j$ and the result strongly depends on
whether these resonances are within the range of integration
$\varepsilon \in\left[0,E_F\right]$. If they are, from the width of
the resonances the amplitude of the exchange interaction energy is
estimated as:
\begin{equation}
E_{res}\sim\sqrt{Tj},
\end{equation}
while the period of the oscillations is
$\hbar/\sqrt{2m\varepsilon_0}$.
 The non-resonant case occurs if $\varepsilon_0\gg E_F$,
$j\ll E_F$. The integration (\ref{eqExchangeFinal}) then results in:
\begin{align}
\label{eqNonRes}  &E_{nr} = \frac{{8\pi T^2 j^2 E_F }}{{\varepsilon
_0 ^4 }}\chi\left(R\right), \nonumber\\
&\chi \left( R \right) = J_0 \left( {k_F R} \right)Y_0 \left( {k_F
R} \right) + J_1 \left( {k_F R} \right)Y_1 \left( {k_F R} \right).
\end{align}
 The condition $j\ll E_F$ allows for the perturbation theory
thus the expression (\ref{eqNonRes}) is what one would expect from
the conventional RKKY approach. The functional dependence on R
$\chi(R)$ is exactly the same as for 2D RKKY interaction without
tunneling\cite{Aristov} and the prefactor accounts for the
particular model we have used to describe the tunneling and the
bound impurity state. The interaction energy amplitude for the
resonance case appears to be substantially higher than for the
non-resonant one. Assuming for both cases $\varepsilon_0\sim E_F$ we
can very roughly estimate the amplification as:
\begin{equation}
\gamma\equiv\frac{{E_{res} }}{{E_{nr} }}\sim\frac{{\varepsilon _0 ^4
}}{{8\pi T^{3/2} j^{3/2} E_F }}.
\end{equation}
For $T\sim0.01 E_F$, $j\sim 0.1 E_F$ $\gamma$ can be as high as 3
orders of magnitude. Fig.\ref{figResonance08Ef} shows the results of
the numerical calculation according to (\ref{eqExchangeFinal}). We
take $m=0.1 m_0$ ($m_0$ -- free electron mass), $E_F=10$ meV,
$\varepsilon_0=0.8 E_F$ so for $j=0.1 E_F$ both integrand resonances
are within the range $[0,E_F]$, for $j=0.3 E_F$ only one resonance
is within the range and the interaction energy is decreased.
The case $j=0.2$ is an intermediate one -- one
of the resonances appears exactly at $E_F$. The non-resonant case is
shown in Fig.\ref{figNonRes}. Here for all the curves both
resonances $\varepsilon=\varepsilon_0\pm j$ are above $E_F$. This
strongly lowers the amplitude of the interaction energy by at least
two orders of magnitude compared to the resonant case in
Fig.\ref{figResonance08Ef}.
 A very different non-resonant limiting case arises from (\ref{eqExchangeFinal})
 for $j\gg E_F,j\gg\varepsilon_0$:
 \begin{equation}
\label{eqNRJ} E_{nrj} = \frac{{8\pi T^2 E_F }}{{j^2 }}\chi \left( R
\right).
\end{equation}
While (\ref{eqNRJ}) has the same dependence on R as
(\ref{eqNonRes}), this it cannot be derived using the perturbation
theory in $j$ and describes the weakening of the interaction at
large $j$ due to the finite energy range of the free carriers
available for the indirect exchange. This case along with the
resonant case may be of importance for the diluted magnetic
semiconductors. For GaAs heterostructures doped with Mn $j$ (unlike
in metals) is commonly assumed to be comparable or even
substantially exceeding $E_F$\cite{Jungwirth}.
In our calculation we have obtained the interaction energy by
analyzing
 the phase shift for
 scattering at the impurity
potential and its effect on the density of states for the standing
waves in a box. This approach, which has been never applied to the
indirect exchange problem before, allowed us to analyze the resonant
case. For the bound state energy being within the energy range
occupied by the free carriers the indirect exchange interaction
appears to be much stronger than expected from the RKKY
approach. We believe that the new results may shed the light on
ferromagnetic coupling in Mn layers in InGaAs-based heterostructures
and other nanostructures with paramagnetic impurities.

The work has been supported by RFBR (grants no 11-02-00348,
11-02-00146, 12-02-00815, 12-02-00141), RF President Grant
NSh-5442.2012.2, Dynasty Foundation.
\bibliography{FanoExchange}
\end{document}